\documentclass[twocolumn,amsmath,amssymb,aps]{revtex4-2}

\usepackage{graphicx}
\usepackage{dcolumn}
\usepackage{bm}
\usepackage{dsfont}
\usepackage{natbib}

\usepackage{graphicx,amsmath}
\usepackage{epsf}

\begin{document}

\title{Entangled Photons and Phonons via Inter-Modal Brillouin Scattering}
\author{Hashem Zoubi}
\email{hashemz@hit.ac.il}
\affiliation{Department of Physics, Holon Institute of Technology, Holon 5810201, Israel}
\date{16 December, 2022}

\begin{abstract}
We explore the possibility of the formation of photon-phonon entangled states in nanoscale wires by exploiting stimulated inter-modal Brillouin scattering of co-propagating photons that belong to distinct spatial optical modes. Inside nanowires, the photon-phonon coupling is significantly enhanced owing to radiation pressure. The Stokes and anti-Stokes processes are decoupled as they involve different phonon modes that lead to symmetry breaking, which results from different phase-matching requirements. For the Stokes process photon-phonon pairs are annihilated or created, in the presence of a classical pump field, and for the anti-Stokes process we obtain coherent oscillations between photons and phonons. The appearance of entangled states can extend the use of nanowires, for example, those made of silicon, into quantum information processing involving photons and phonons in a setup that can be easily integrated into an on-chip network.
\end{abstract}

\maketitle

\section{Introduction}

Light propagates in the forward direction inside a homogeneous isotropic dielectric medium and scatters due to fluctuations in the optical properties. Stimulated Brillouin's Scattering (SBS) is a significant inelastic scattering of light from sound waves in the medium. The mechanism behind SBS is electrostriction in which sound waves are induced by the presence of light and that results in light scattering \cite{Boyd2008}. In one-dimensional systems (dielectric waveguides, e.g. optical fibers made of silica), the SBS is significantly enhanced linearly by the fiber length and inversely by the fiber cross-section area, where the backward SBS dominates over the forward one \cite{Kobyakov2010,Agrawal2013}. SBS in waveguides is involved in a number of useful applications for photonics and fruitful information processing. For example, SBS can be tailored for realizing slow light, storage of light, narrow linewidth laser sources, optical frequency comb generation, high performance signal processing and sensing \cite{Eggleton2013}.

Big effort has been devoted for fabricating waveguides with smaller and smaller cross section \cite{Safavi2019}. A breakthrough has been achieved when the waveguide dimension approaches the nanoscale regime \cite{Rakich2012,Sipe2016,Zoubi2016}. Hence a new mechanism of radiation pressure is started to participate and enhanced the SBS strength by several orders of magnitudes \cite{Shin2013,Kittlaus2015,VanLaer2015a,VanLaer2015b,Huy2016}. Radiation pressure dominates over electrostriction as the cross-section dimension becomes smaller than the light wavelength, where the light apply forces on the waveguide boundaries and that induces additional vibrational modes which significantly increases SBS. Nanowires made of silicon are among the successful candidates for obtaining strong SBS and they are of importance for achieving all-optical devices that can be easily integrated into on-chip platforms \cite{Eggleton2019}. Moreover, forward and backward SBS take place in nanoscale waveguides which is important in the present paper. A wide range of useful applications appears once we enter the nanoscale regime, e.g. in quantum communication, quantum sensing and quantum information processing. Coherent phenomena in nanoscale waveguides have been demonstrated experimentally, e.g. for silicon Brillouin laser \cite{Otterstrom2017}, photonic-phononic memory devices \cite{Merklein2017}, and side-band cooling \cite{Otterstrom2018}, and suggested theoretically, e.g. for quantum logic gates and non-classical states \cite{Zoubi2017,Zoubi2018,Zoubi2019,Zoubi2020,Zoubi2021}.

Photons and phonons are strongly coupled in optomechanical nanophotonic waveguides \cite{Kittlaus2015,VanLaer2015a}. The photon and phonon confinement within nanophotonic waveguides lead to significant Brillouin nonlinear interactions. Strong SBS of multi-mode light-fields belong to distinct spatial modes have been demonstrated in on-chip integrated silicon waveguides that termed stimulated inter-modal Brillouin scattering \cite{Kittlaus2017}. This inter-modal SBS decouples Stokes and anti-Stokes processes. We consider forward SBS within silicon waveguides between co-propagating light fields guided in distinct spatial modes that enable only single-sideband effect. The phase-matching requirement in inter-modal Brillouin scattering results in symmetry breaking between Stokes and anti-Stokes processes by coupling them to different phonon modes.

Strong forward SBS have been demonstrated within silicon waveguides, where the coupling between co-propagating light fields that are guided in the same optical mode is mediated by phonons \cite{kharel2016,Wolff2017}. Forward SBS between photons within the same optical mode produces two-sideband effect, the fact that limit the performance of such processes. Here both Stokes and anti-Stokes scattering processes are mediated by the same phonon mode and as a result yields symmetric scattering of photons with higher and lower frequencies. On the other hand, integrated optomechanical waveguides give rise to multi-mode light fields guided in distinct spatial modes \cite{Kittlaus2017}.  Forward SBS through the coupling between photons guided in distinct spatial modes mediated by phonons, termed inter-modal SBS, enabling single-sideband effect by decoupling Stokes and anti-Stokes processes. Here the phase-matching requirement produces a symmetry breaking that causes the Stokes and anti-Stokes processes to couple to different phonon modes.

In the present paper we study the possibility of the formation of photon and phonon entangled states in optomechanical nanophotonic silicon waveguides by exploiting Brillouin interactions which result from coherent coupling between controlled photons and phonons. As for inter-modal SBS the system decouples Stokes and anti-Stokes processes we consider the two cases separately and investigate the appearance of photon-phonon entangled states, which are the cornerstone for quantum information processing and quantum communications. The quantum Brillouin Hamiltonian is nonlinear, as the interaction term includes three operators, two photonic and one phononic \cite{Zoubi2016}. In the Stokes process a pump photon is scattered into a Stokes photon and the emission of a Stokes phonon, while for anti-Stokes process a pump photon is scattered into an anti-Stokes photon and the absorption of an anti-Stokes phonon. The two processes obey conservation of energy and momentum, and even though the Stokes and anti-Stokes phonons have the same frequency they differ in their wavenumbers. The Hamiltonian can be linearized by taking the strong pump field to be a classical one. We get two separated linear interaction Hamiltonians, the Stokes Hamiltonian and the anti-Stokes Hamiltonian, the fact that allows us to treat the two processes separately. The diagonalization of the linear Hamiltonian leads to new diagonal quantum states which are a coherent mix of photons and phonons (in the presence of the strong pump field). The diagonal states are obtained by applying the diagonal operators to the vacuum state. At the Stokes process the diagonal state represents creation or annihilation of photon-phonon pairs, while at the anti-Stokes process the diagonal state represents coherent oscillations between photons and phonons. We investigate the properties of such states and emphasize their importance for quantum information processing.

\section{Stimulated Inter-modal Forward Brillouin Scattering}

We consider a one dimensional waveguide build of a nanowire made of silicon \cite{Kittlaus2017}. The nanowire is interfaced with two integrated multiplexers ($M_r$ and $M_l$) that allow internal-external coupling of optical modes, as appears in figure (1.a). The system supports the propagation of hybrid photonic and phononic multi-mode \cite{Rakich2012,Zoubi2016}. Two of the optical spatial branches are presented in figure (1.b) and one of the mechanical vibrational branches is presented in figure (1.c). The photon dispersion is taken to be linear at the appropriate region, where $\omega_{ik}=\omega_{i0}+v_gk$. Here $(i=1,2)$ stands for the two branches and $\omega_{i0}$ is a given frequency for each branch. The photons at the two branches have the same group velocity $v_g$, with the photon wavenumber $k$, where we consider photons propagating only to the right. The photon Hamiltonian of the two branches reads
\begin{equation}
\hat{H}_{phot}=\sum_i\sum_{k}\hbar\omega_{ik}\ \hat{a}_{ik}^{\dagger}\hat{a}_{ik},
\end{equation}
where $\hat{a}_{ik}^{\dagger}$ and $\hat{a}_{ik}$ are the creation and annihilation operators of wavenumber $k$ at branch $i$. The phonon Hamiltonian reads
\begin{equation}
\hat{H}_{phon}=\sum_{q}\hbar\Omega_{q}\ \hat{b}_{q}^{\dagger}\hat{b}_{q},
\end{equation}
where $\hat{b}_{q}^{\dagger}$ and $\hat{b}_{q}$ are the creation and annihilation operators of wavenumber $q$. The vibrational phonon is dispersionless of frequency $\Omega_q=\Omega$. The phonons can propagate to the left or the right. Note that $\omega_{i0}$ and the finite $\Omega$ appear due to the photon and phonon confinement within the nanoscale waveguide. The inter-modal SBS Hamiltonian is given by
\begin{equation}
\hat{H}_{I}=\hbar\sum_{ij}\sum_{kq}\left(g^{\ast}_{kq}\ \hat{b}_q^{\dagger}\hat{a}_{jk-q}^{\dagger}\hat{a}_{ik}+g_{kq}\ \hat{b}_q\hat{a}_{ik}^{\dagger}\hat{a}_{jk-q}\right),
\end{equation}
where $g_{kq}$ is the SBS coupling parameter between two photons with wavenumbers $k$ and $k-q$ and a phonon of wavenumber $q$. The photons belong to two different branches where $(i\neq j)$ that are subjected to the conservation of momentum. In the following we treat the Stokes and anti-Stokes processes separately. The light fields are propagating to the right and can be excited through the left multiplexer. The input and output fields at wavenumber $k$ is are represented by the operators $\hat{c}_k^{in}$ and $\hat{c}_k^{out}$. From the input-output formalism we have the boundary condition $\hat{c}_k^{in}+\hat{c}_k^{out}=\sqrt{u_k}\hat{a}_k$, where $u_k$ is the coupling parameter at the multiplexer between the external and the internal fields \cite{Gardiner2010,Walls2008}.

\begin{figure}
\includegraphics[width=1.0\linewidth]{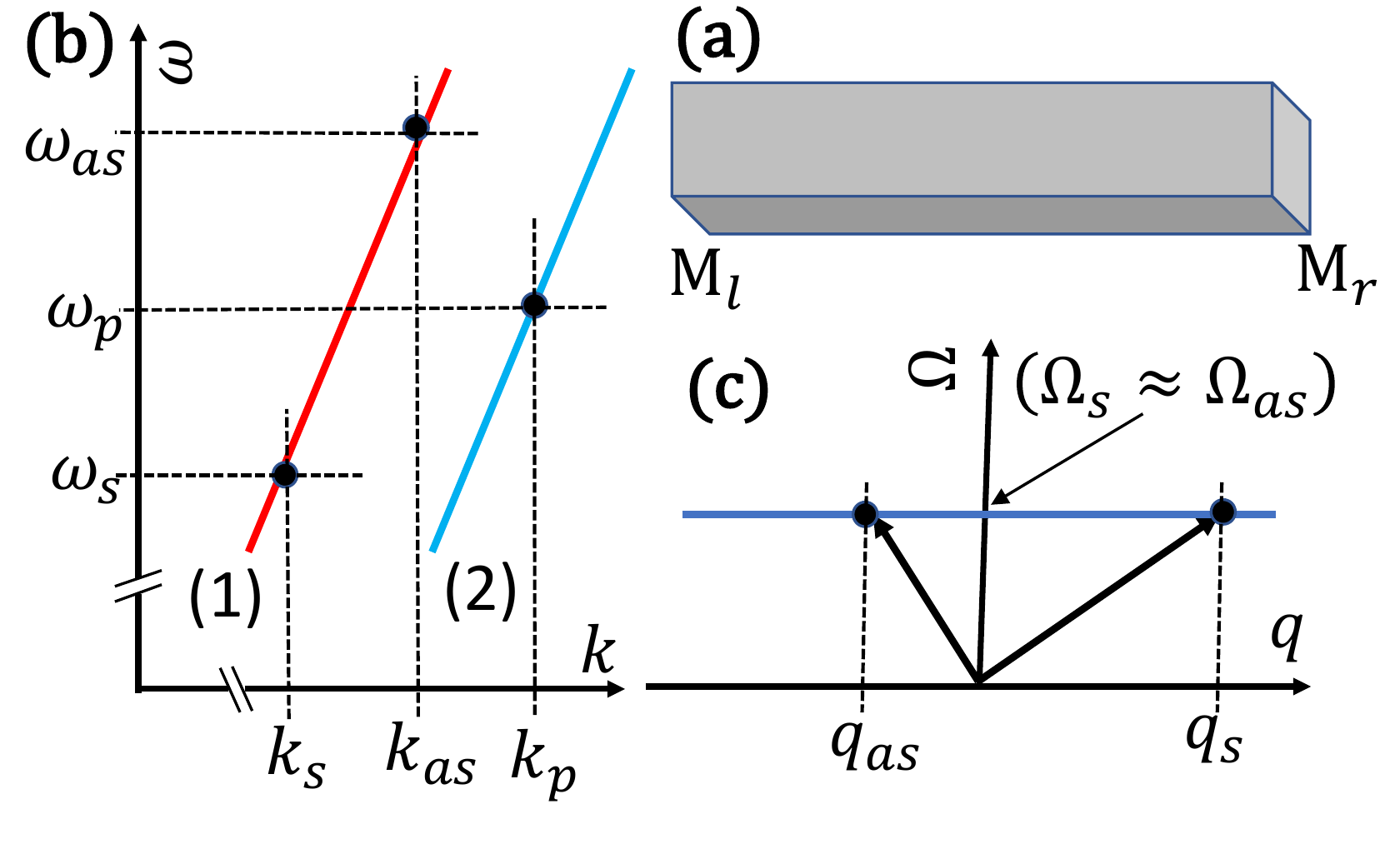}
\caption{(a) A nanowire is presented that interfaced by two multiplexers $M_r$ and $M_l$. (b) Two optical spatial branches of linear dispersion are plotted for the frequency $\omega$ as a function of the wavenumber $k$. The three points of importance in the paper are indicated, the pump field $(\omega_p,k_p)$, the Stokes field $(\omega_s,k_s)$, and the anti-Stokes field $(\omega_{as},k_{as})$. (c) The vibrational phonon is presented for the frequency $\Omega$ as a function of the wavenumber $q$. The Stokes and anti-Stokes phonons of $(\Omega_{s},q_{s})$ and $(\Omega_{as},q_{as})$ are indicated. As the phonon bransh is dispersionless we have $\Omega_s=\Omega_{as}$.}
\end{figure}

Photons can leak out of the nanowire with damping rate $\gamma$. Usually in nanowires the photon damping rate is much smaller than the photon-phonon coupling, that is $\gamma\ll |g|$. The average number of thermal photons of optical frequencies is much smaller than "one" at room temperature, and hence can be neglected. Phonon damping could have important influence for the discussion of the present paper. The damping rate of phonons, $\Gamma$ ,is of the order of the photon-phonon coupling, that is $\Gamma\sim|g|$ \cite{Kharel2016,VanLaer2017}. We show later, after linearizing the Hamiltonian, how one can achieve the limit of phonon damping rate much smaller than the effective photon-phonon coupling for both cases of Stokes and anti-Stokes processes. The average number of thermal phonons at frequency of about 10 GHz, which is of interest in the current paper, is much larger than "one" at room temperature. Therefore, in order to decrease the average number of thermal phonons one needs to cool the nanowire. Hence, to reach average number of thermal phonons much smaller than  "one" the temperature needs to be around 10 mK.

\subsection{Stokes Process}

For the Stokes process a pump photon of wavenumber $k_p$ and frequency $\omega_p$ at branch $2$ scatters to a Stokes photon of wavenumber $k_s$ and frequency $\omega_s$ at branch $1$ and a phonon of wavenumber $q_s$ and frequency $\Omega_s$, as depicted in figure (2). Conservation of energy implies $\omega_p\approx\omega_s+\Omega_s$, where $\omega_p>\omega_s$. The conservation of momentum implies the phase matching $k_p=k_s+q_s$. Note that the two photons and the phonon are propagating in the same direction as presented in figure (3). Once, in the presence of the pump field, a probe field at the Stokes frequency is sent into the waveguide, the Stokes process becomes dominant and the other processes are suppressed. The Stokes process Hamiltonian is represented by
\begin{eqnarray}
\hat{H}_S&=&\hbar\omega_p\ \hat{a}_p^{\dagger}\hat{a}_p+\hbar\omega_s\ \hat{a}_s^{\dagger}\hat{a}_s+\hbar\Omega_s\ \hat{b}_s^{\dagger}\hat{b}_s \nonumber \\
&+&\hbar g_s^{\ast}\ \hat{b}_s^{\dagger}\hat{a}_s^{\dagger}\hat{a}_p+\hbar g_s\ \hat{b}_s \hat{a}_s \hat{a}_p^{\dagger},
\end{eqnarray}
where $\hat{a}_p^{\dagger}$ and $\hat{a}_p$ are the pump photon operators,  $\hat{a}_s^{\dagger}$ and $\hat{a}_s$ are the Stokes photon operators, $\hat{b}_s^{\dagger}$ and $\hat{b}_s$ are the Stokes phonon operators. The photon-phonon coupling parameter is $g_s$ for the Stokes process. Here a pump photon is absorbed and a Stokes photon with a Stokes phonon are emitted, and the opposite a Stokes photon with a Stokes phonon are absorbed and a pump photon is emitted.

The Stokes process Hamiltonian is nonlinear due to the three-operator term. The Hamiltonian can be linearized by converting the pump field operator into a classical field which is allowed at strong pump field. The pump field is excited through an external field that described by $\hat{c}_p^{in}$ at frequency $\omega_p$. The input-output formalism yields the equation of motion for the pump operator $\frac{d}{dt}\tilde{a}_p\approx-u\tilde{a}_p+\sqrt{u}\tilde{c}_p^{in}$, where we defined $\tilde{a}_p=\hat{a}_p e^{i\omega_pt}$ and $\tilde{c}_p^{in}=\hat{c}_p^{in}e^{i\omega_pt}$, and $u$ is the external-internal coupling parameter at the multiplexer. At steady state we have $\frac{d}{dt}\tilde{a}_p=0$, hence we get $\hat{a}_p=\frac{1}{\sqrt{u}}\hat{c}_p^{in}$. The number operator is defined by $\hat{n}_p=\hat{a}_p^{\dagger}\hat{a}_p=\frac{1}{u}\hat{c}_p^{in\dagger}\hat{c}_p^{in\dagger}$, and the expectation value yields $n_p=\langle\hat{a}_p^{\dagger}\hat{a}_p\rangle$ and $n_p^{in}=\langle \hat{c}_p^{in\dagger}\hat{c}_p^{in\dagger}\rangle$, then we have $n_p=\frac{n_p^{in}}{u}$. We replace the pump photon operator by its average value where $\langle \hat{a}_p\rangle=\sqrt{n_p}=\sqrt{\frac{n_p^{in}}{u}}$. We obtain the Stokes process Hamiltonian
\begin{eqnarray} \label{HS}
\hat{H}_S&=&\hbar\omega_s\ \hat{a}_s^{\dagger}\hat{a}_s+\hbar\Omega_s\ \hat{b}_s^{\dagger}\hat{b}_s \nonumber \\
&+&\hbar\sqrt{\frac{n_p^{in}}{u}}\left(g_s^{\ast}\ \hat{b}_s^{\dagger}\hat{a}_s^{\dagger}+\hbar g_s\ \hat{b}_s \hat{a}_s \right).
\end{eqnarray}

\begin{figure}
\includegraphics[width=0.6\linewidth]{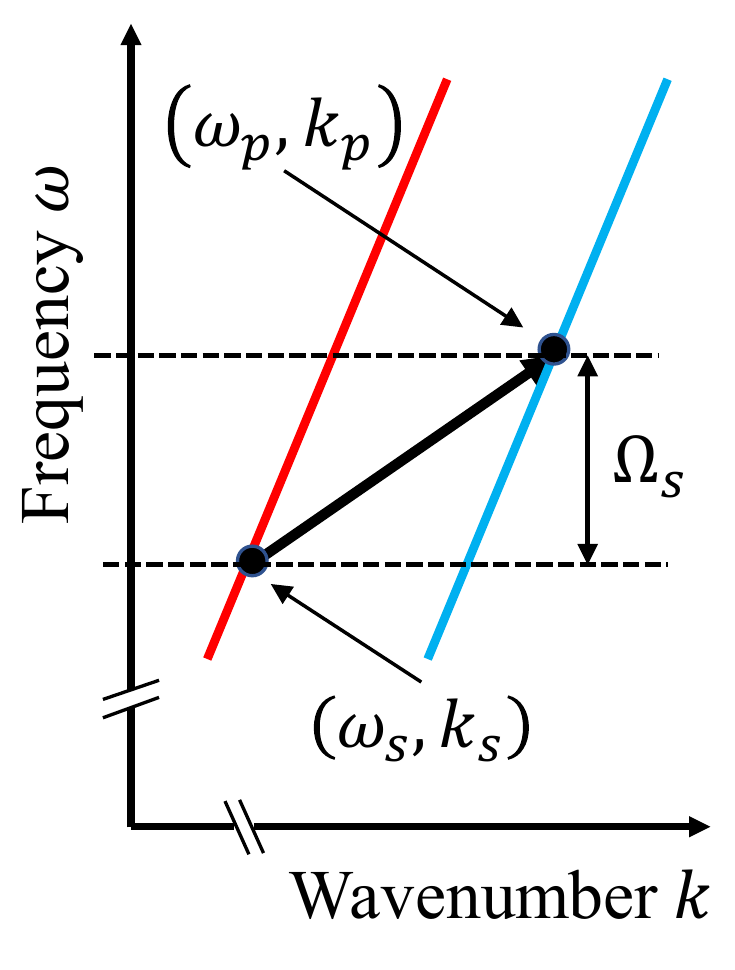}
\caption{The Stokes process is presented, where a pump photon $(\omega_p,k_p)$ is scattered into a Stokes photon $(\omega_s,k_s)$ and a Stokes phonon $(\Omega_s,q_s)$.}
\end{figure}

\begin{figure}
\includegraphics[width=0.8\linewidth]{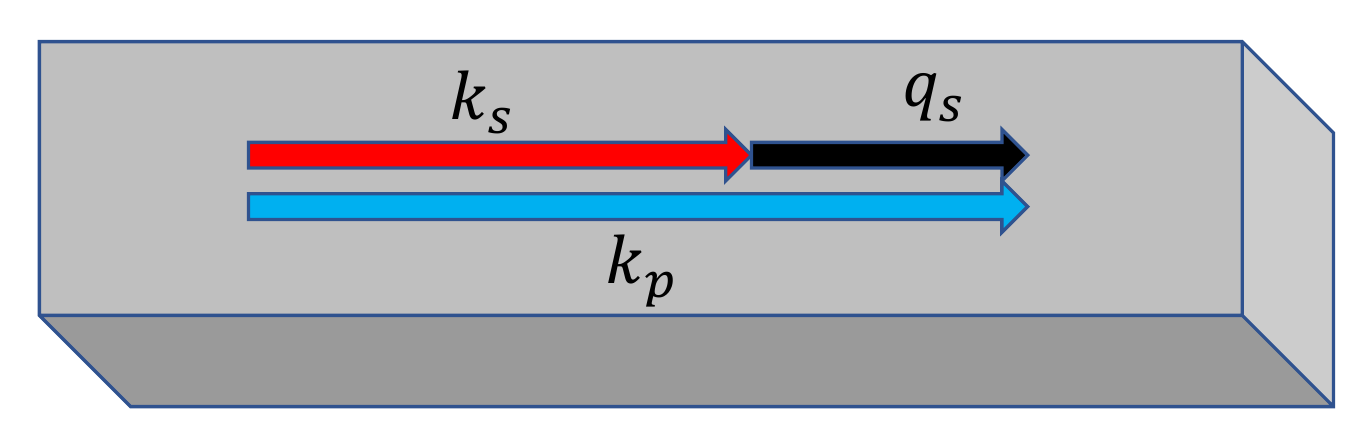}
\caption{Schematic presentation of the Stokes process inside a nanowire. The pump photon of wavenumber $k_p$ is scattered into a Stokes photon of $k_s$ and a Stokes phonon of $q_s$, which are subjected to the conservation of momentum where $k_p=k_s+q_s$.}
\end{figure}

\subsection{Anti-Stokes Process}

For the anti-Stokes process a pump photon of wavenumber $k_p$ and frequency $\omega_p$ at branch $2$ scatters to an anti-Stokes photon of wavenumber $k_{as}$ and frequency $\omega_{as}$ at branch $1$ with a phonon of wavenumber $q_{as}$ and frequency $\Omega_{as}$, as depicted in figure (4). Conservation of energy implies $\omega_p\approx\omega_{as}-\Omega_{as}$, where $\omega_p<\omega_{as}$. The conservation of momentum implies the phase matching $k_p=k_{as}+q_{as}$. Note that the two photons are propagating in the same direction and the phonon in the opposite one as presented in figure (5). Once, beside the pump field, a probe field at the anti-Stokes frequency is sent into the waveguide, the anti-Stokes process becomes dominant and the other processes are negligible. The anti-Stokes process Hamiltonian is represented by
\begin{eqnarray}
\hat{H}_{AS}&=&\hbar\omega_p\ \hat{a}_p^{\dagger}\hat{a}_p+\hbar\omega_{as}\ \hat{a}_{as}^{\dagger}\hat{a}_{as}+\hbar\Omega_{as}\ \hat{b}_{as}^{\dagger}\hat{b}_{as} \nonumber \\
&+&\hbar g_{as}\ \hat{a}_{as}^{\dagger}\hat{a}_p \hat{b}_{as}+\hbar g_{as}^{\ast}\ \hat{a}_{as} \hat{a}_p^{\dagger}\hat{b}_{as}^{\dagger},
\end{eqnarray}
where $\hat{a}_{as}^{\dagger}$ and $\hat{a}_{as}$ are the anti-Stokes photon operators, $\hat{b}_{as}^{\dagger}$ and $\hat{b}_{as}$ are the anti-Stokes phonon operators. The photon-phonon coupling parameter is $g_{as}$ for the anti-Stokes process. Here a pump photon and a phonon are absorbed and an anti-Stokes photon is emitted, and the opposite an anti-Stokes photon is absorbed and a pump photon and a phonon are emitted.

\begin{figure}
\includegraphics[width=0.7\linewidth]{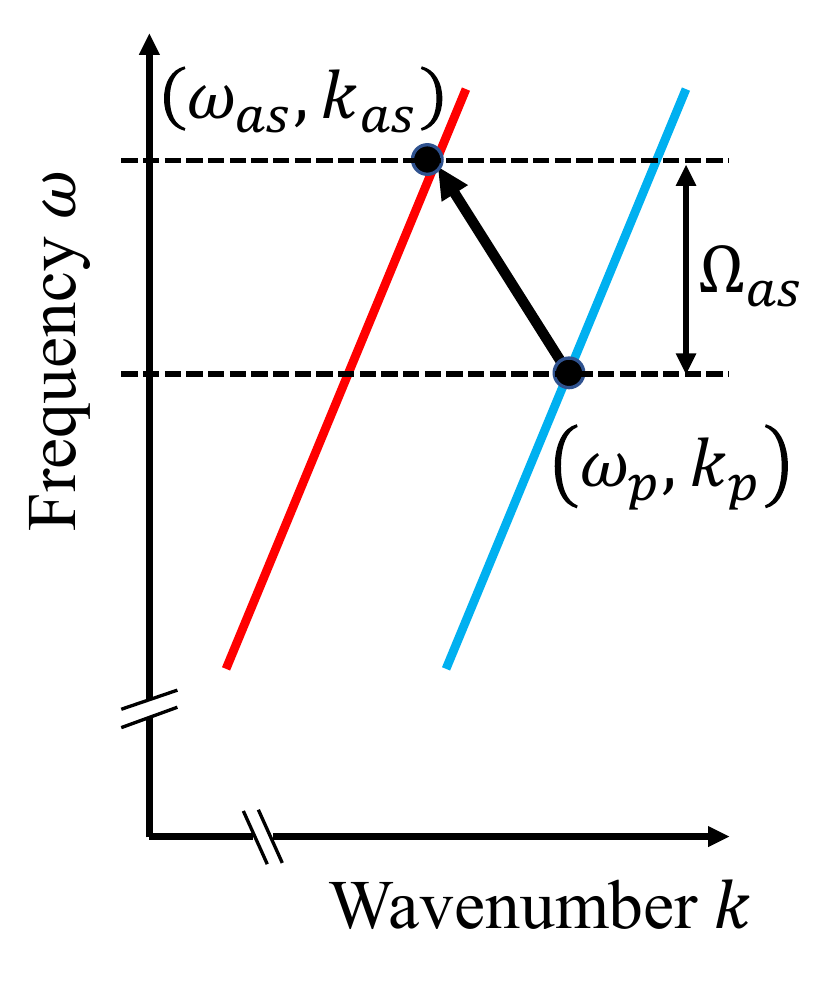}
\caption{The anti-Stokes process is presented, where a pump photon $(\omega_p,k_p)$ is scattered into an anti-Stokes photon $(\omega_{as},k_{as})$ by the absorption of an anti-Stokes phonon $(\Omega_{as},q_{as})$.}
\end{figure}

\begin{figure}
\includegraphics[width=0.8\linewidth]{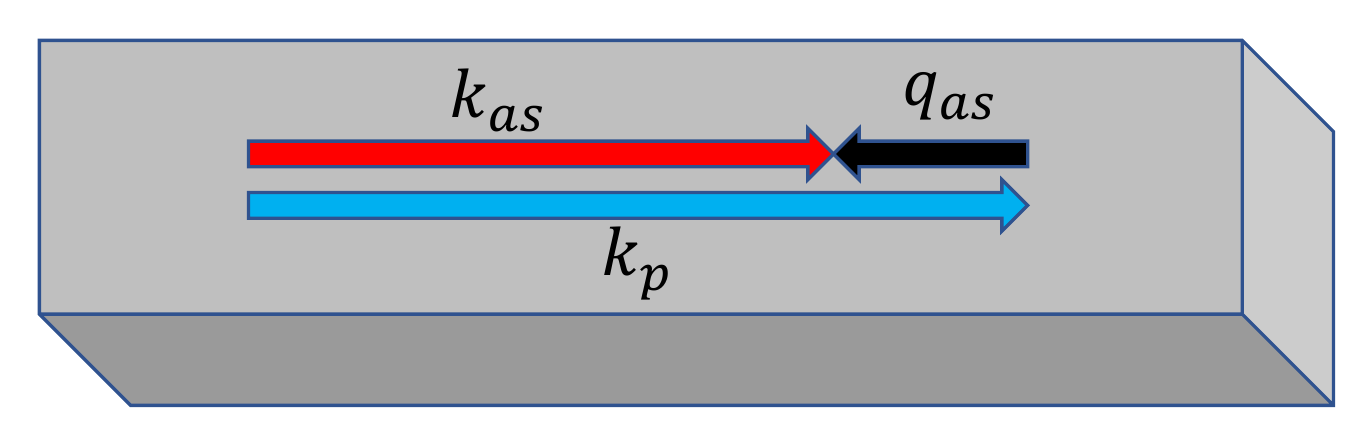}
\caption{Schematic presentation of the anti-Stokes process inside a nanowire. The pump photon of wavenumber $k_p$ is scattered into an anti-Stokes photon of $k_{as}$ by the absorption of an anti-Stokes phonon of $q_{as}$, which are subjected to the conservation of momentum where $k_p=k_{as}-(-q_{as})$.}
\end{figure}

The anti-stokes process Hamiltonian is nonlinear due to the three-operator term, and as before, can be linearized by converting the pump field operator into a classical field. We obtain the anti-Stokes process Hamiltonian
\begin{eqnarray}\label{HAS}
\hat{H}_{AS}&=&\hbar\omega_{as}\ \hat{a}_{as}^{\dagger}\hat{a}_{as}+\hbar\Omega_{as}\ \hat{b}_{as}^{\dagger}\hat{b}_{as} \nonumber \\
&+&\hbar\sqrt{\frac{n_p^{in}}{u}}\left(g_{as}^{\ast}\ \hat{b}_{as}^{\dagger}\hat{a}_{as}+\hbar g_{as}\ \hat{b}_{as} \hat{a}_{as}^{\dagger} \right).
\end{eqnarray}

\section{Photon and phonon entangled states}

In the next step we aim to diagonalize the Hamiltonians in a rotating frame. We treat the two cases of Stokes and anti-Stokes processes separately. The diagonalization is obtained in defining collective operators. The application of the collective operators on the ground state gives rise to photon-phonon entangled states. Then we examine the properties of such states. Note that the difference between the two phonons, which is due to the difference in their wavenumbers, $q_s\neq q_{as}$, even though they have the same frequency $\Omega_s=\Omega_{as}$, allows one to discriminate between the Stoked and the anti-Stokes processes.

\subsection{Entangled states via Stokes processes}

The Stokes process Hamiltonian (\ref{HS}) in a rotating frame that oscillates with the pump frequency is given by the quadratic Hamiltonian
\begin{equation}
\hat{H}_S=\hbar\Delta\omega_s\ \hat{a}_s^{\dagger}\hat{a}_s+\hbar\Omega_s\ \hat{b}_s^{\dagger}\hat{b}_s+\hbar f_s\left(\hat{b}_s^{\dagger}\hat{a}_s^{\dagger}+\hat{b}_s \hat{a}_s\right),
\end{equation}
where $\Delta\omega_s=\omega_p-\omega_s$, and $f_s=g_s\sqrt{\frac{n_p^{in}}{u}}$ that is taken to be real. The Hamiltonian can be easily diagonalized using the Bogoliubov transformation \cite{Fetter1971}
\begin{eqnarray}\label{TRANS}
\hat{\alpha}&=&\cosh r\ \hat{a}_s+\sinh r\ \hat{b}_s^{\dagger}, \nonumber \\
\hat{\beta}&=&\cosh r\ \hat{b}_s+\sinh r\ \hat{a}_s^{\dagger},
\end{eqnarray}
with the inverse transformation
\begin{eqnarray}
\hat{a}_s&=&\cosh r\ \hat{\alpha}-\sinh r\ \hat{\beta}^{\dagger}, \nonumber \\
\hat{b}_s&=&\cosh r\ \hat{\beta}-\sinh r\ \hat{\alpha}^{\dagger}.
\end{eqnarray}
The fact that $\hat{a}_s$ and $\hat{b}_s$ obey boson commutation relations, $[\hat{a}_s,\hat{a}_s^{\dagger}]=[\hat{b}_s,\hat{b}_s^{\dagger}]=1$ and $[\hat{a}_s,\hat{b}_s^{\dagger}]=0$, lead to $[\hat{\alpha},\hat{\alpha}^{\dagger}]=[\hat{\beta},\hat{\beta}^{\dagger}]=1$ and $[\hat{\alpha},\hat{\beta}^{\dagger}]=0$, hence $\cosh^2 r-\sinh^2 r=1$. Substitution in the Hamiltonian gives
\begin{eqnarray}
\hat{H}_S&=&\hbar\left[(\Delta\omega_s+\Omega_s)\sinh^2 r-2f_s\cosh r\sinh r\right]\hat{\mathbb{I}} \nonumber \\
&+&\hbar\left[\Delta\omega_s\cosh^2 r+\Omega_s\sinh^2 r-2f_s\cosh r\sinh r\right]\hat{\alpha}^{\dagger}\hat{\alpha} \nonumber \\
&+&\hbar\left[\Omega_s\cosh^2 r+\Delta\omega_s\sinh^2 r-2f_s\cosh r\sinh r\right]\hat{\beta}^{\dagger}\hat{\beta} \nonumber \\
&-&\hbar\left[(\Delta\omega_s+\Omega_s)\cosh r\sinh r-f_s(\cosh^2 r+\sinh^2 r)\right] \nonumber \\
&\times&\left(\hat{\alpha}\hat{\beta}+\hat{\alpha}^{\dagger}\hat{\beta}^{\dagger}\right),
\end{eqnarray}
where $\hat{\mathbb{I}}$ is a unit operator. We choose $(\Delta\omega_s+\Omega_s)\cosh r\sinh r=f_s(\cosh^2 r+\sinh^2 r)$, to get the diagonal Hamiltonian
\begin{equation}
\hat{H}_S=\hbar\omega_0\ \hat{\mathbb{I}}+\hbar\omega_{\alpha}\ \hat{\alpha}^{\dagger}\hat{\alpha}+\hbar\omega_{\beta}\ \hat{\beta}^{\dagger}\hat{\beta},
\end{equation}
where
\begin{eqnarray}
\omega_0&=&(\Delta\omega_s+\Omega_s)\sinh^2 r-2f_s\cosh r\sinh r, \nonumber \\
\omega_{\alpha}&=&\Delta\omega_s\cosh^2 r+\Omega_s\sinh^2 r-2f_s\cosh r\sinh r, \nonumber \\
\omega_{\beta}&=&\Omega_s\cosh^2 r+\Delta\omega_s\sinh^2 r-2f_s\cosh r\sinh r.
\end{eqnarray}
The calculation yields
\begin{equation}
\cosh^2 r=\frac{\bar{\omega}+\Delta_s}{2\Delta_s},\ \sinh^2 r=\frac{\bar{\omega}-\Delta_s}{2\Delta_s},
\end{equation}
and $\cosh r\sinh r=\frac{f_s}{2\Delta_s}$, where we defined $\Delta_s^2=\bar{\omega}^2-f_s^2$, with $\bar{\omega}=\frac{\Delta\omega_s+\Omega_s}{2}$. We obtain
\begin{equation}
\omega_{\alpha}=\Delta_s+\delta_s,\ \omega_{\beta}=\Delta_s-\delta_s,
\end{equation}
and $\omega_0=\Delta_s-\bar{\omega}$,
where $\delta_s=\frac{\Delta\omega_s-\Omega_s}{2}$.

The ground state $|vac\rangle$ has energy $\hbar\omega_0$, and we get two collective Bogoliubov modes with energies $\hbar\omega_{\alpha}$ and $\hbar\omega_{\beta}$, which are represented by the boson operators $\hat{\alpha}$ and $\hat{\beta}$, respectively. The creation and annihilation of a collective excitation involves creation and annihilation of a photon-phonon pair in the presence of an external pump field. The collective states represent photon-phonon entangled states. The stability requirement of such entangled states implies the condition $\bar{\omega}>f_s$.

For typical silicon nanowires we use the following numbers \cite{Kittlaus2015,VanLaer2015a}. The photon-phonon coupling parameter, $g_s$, is $1$ MHz, and for the internal-external coupling parameter, $u$, at the multiplexers we have $1$ MHz, and the average number of the incident pump photons is taken to be of the order of $10^{12}$ per second. Hence for the effective photon-phonon coupling parameter we get $f_s=1$ GHz. For the pump photon we use $\omega_p=10^{15}$ Hz, and for the phonon we have $\Omega_s=10$ GHz. The diagonal frequencies are plotted in figures (6) and (7) as a function of the detuning $\delta_s$. Here $\omega_{\alpha}$ grows linearly with  $\delta_s$ while  $\omega_{\beta}$ changes slightly and the ground frequency $\omega_0$ stays small, $\omega_0\ll\Omega_s$. The intersection point of $\omega_{\alpha}=\omega_{\beta}$ appears at $\delta_s=0$. The fractions, $\cosh^2 r$ and $\sinh^2 r$, as a function of $\delta_s$ are plotted in figure (8), and in much details in figure (9). Here $\cosh^2 r\approx 1$ is around one, and $\sinh^2 r\approx 0$ is around zero.

The phonon damping rate is about $\Gamma=1$ MHz \cite{Kharel2016}, which is of the order of the photon-phonon coupling parameter, $g_s$. But, as the effective photon-phonon coupling parameter $f_s$ is much larger than the phonon damping rate, we are in the strong coupling regime, and hence the appearance of the collective Bogoliubov modes is feasible. On the other hand, thermal phonons of 10 GHz frequency are dominant at room temperature. Therefore in order to minimize the influence of the thermal phonons one needs to cool the system down to 10 mK.

\begin{figure}
\includegraphics[width=0.8\linewidth]{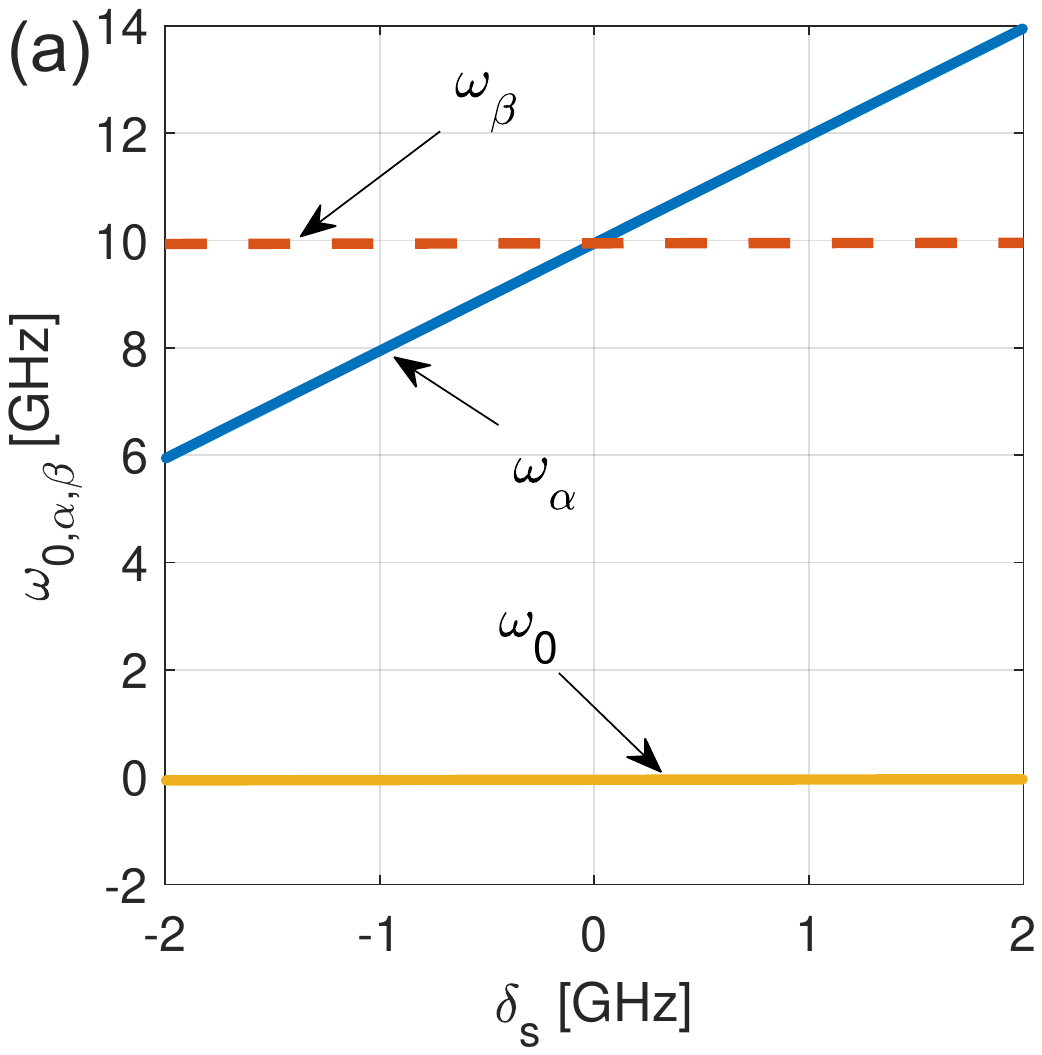}
\includegraphics[width=0.8\linewidth]{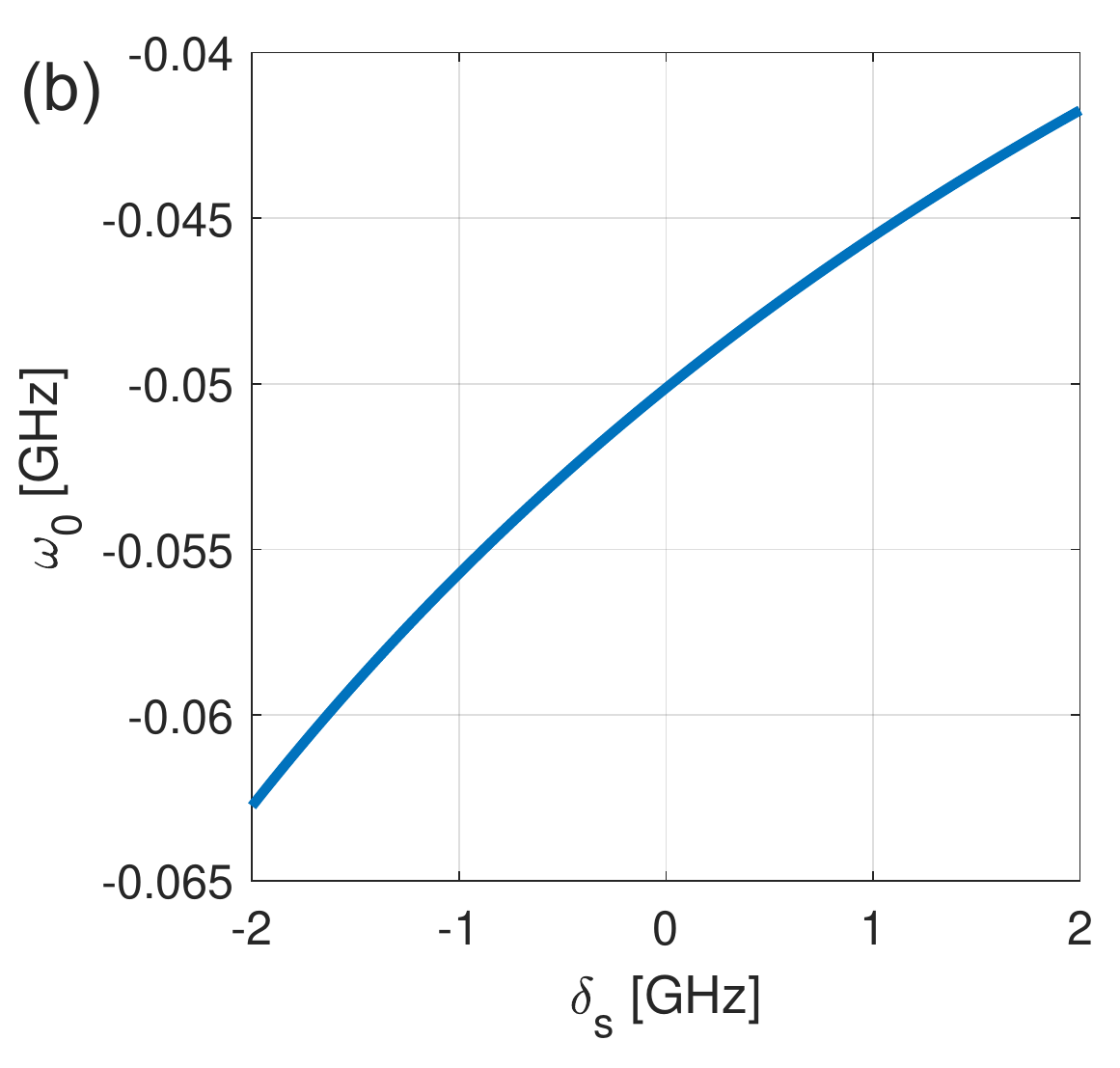}
\caption{(a) The two diagonal Bogoliubov modes $\omega_{\alpha}$ and $\omega_{\beta}$ and the ground state $\omega_0$ are presented as a function of the detuning $\delta_s$. (b) The ground state $\omega_0$ is plotted as a function of $\delta_s$, where the frequency is slightly changed relative to the Bogoliubov modes and stays close to zero, $\omega_0\approx 0$.}
\end{figure}

\begin{figure}
\includegraphics[width=0.8\linewidth]{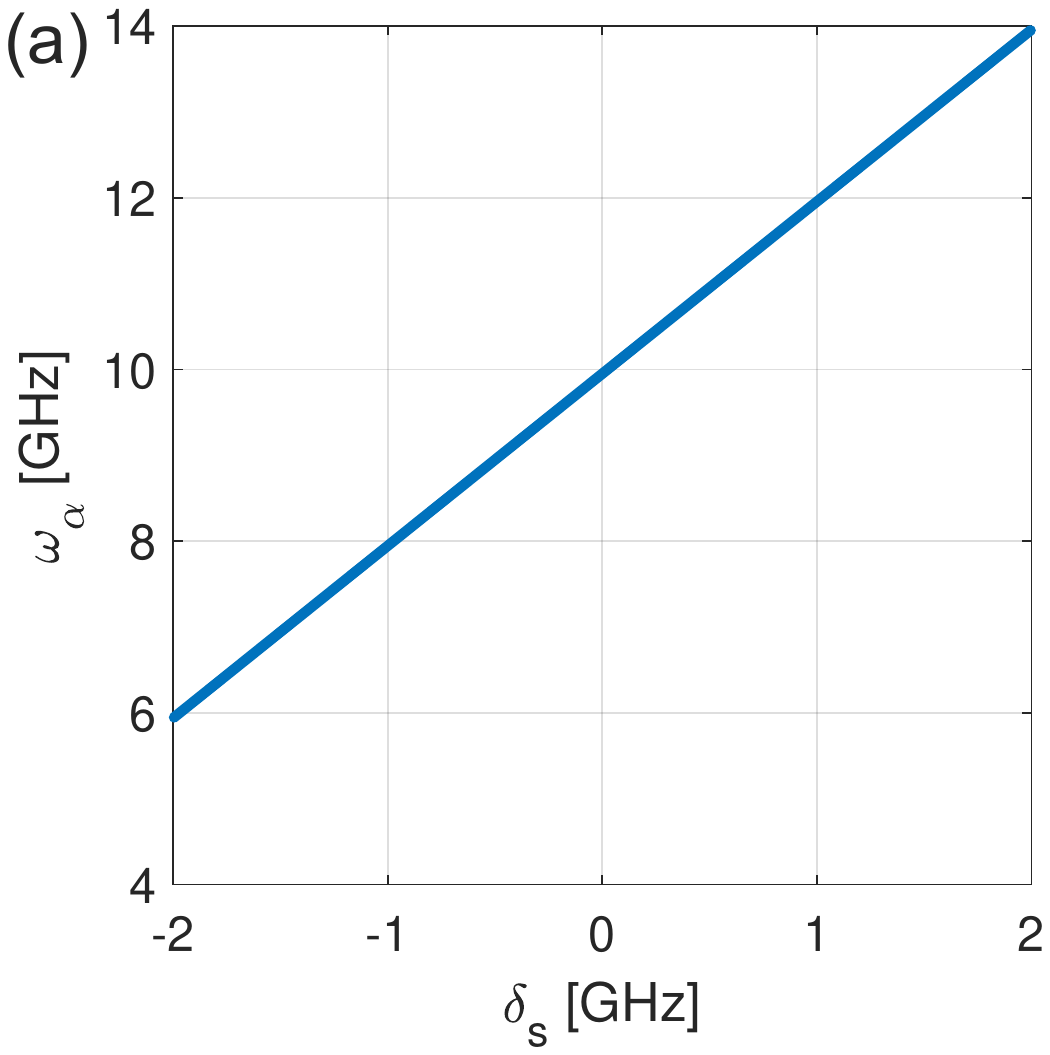}
\includegraphics[width=0.8\linewidth]{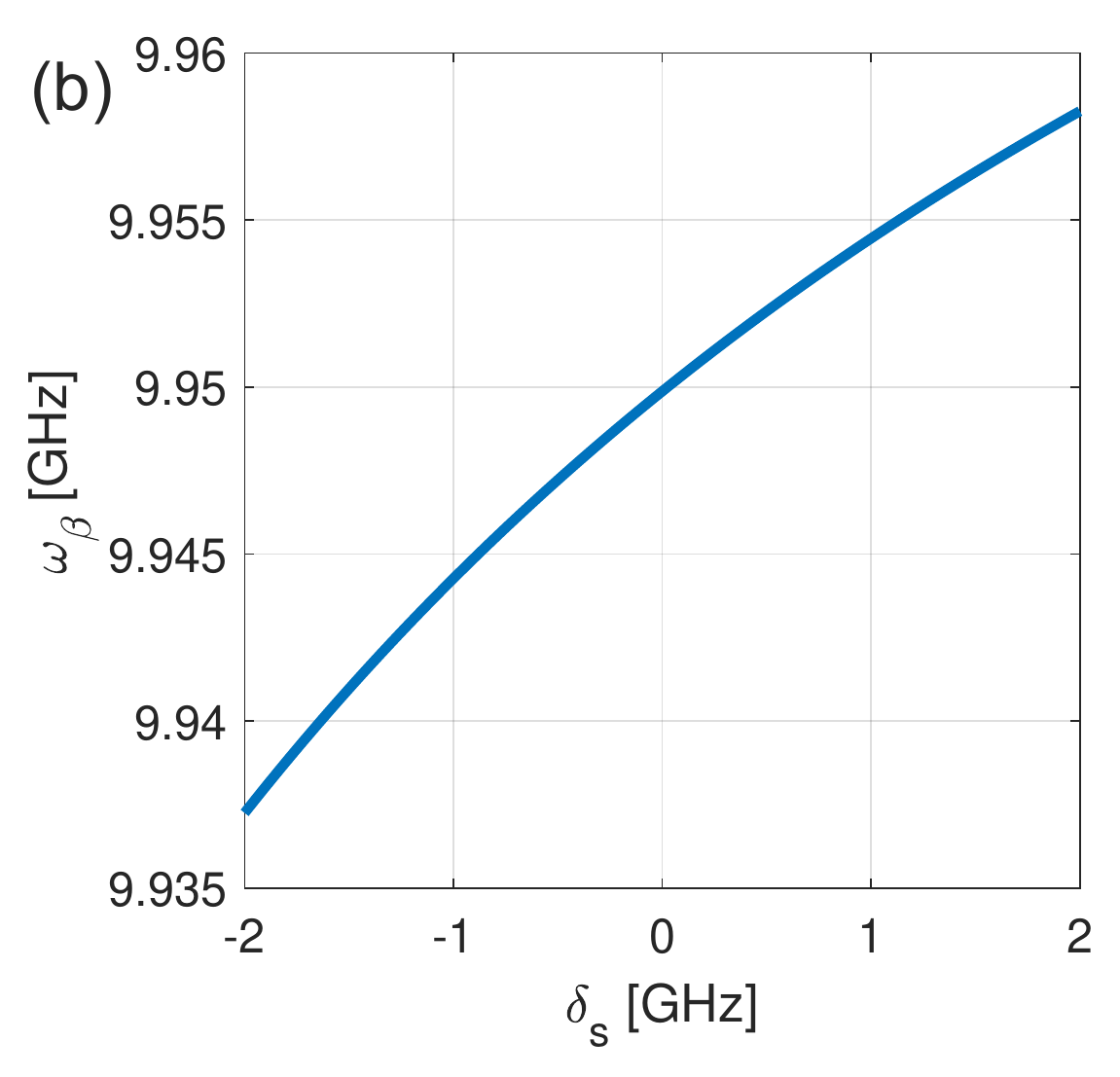}
\caption{(a) The Bogoliubov mode $\omega_{\alpha}$ is plotted as a function of the detuning $\delta_s$. The mode shows linear behavior around $\delta_0\approx 0$. (b) The Bogoliubov mode $\omega_{\beta}$ is plotted as a function of the detuning $\delta_s$. The mode is almost constant and slightly changes from the phonon frequency, that is $\omega_{\alpha}\approx \Omega_s$.}
\end{figure}

\begin{figure}
\includegraphics[width=0.8\linewidth]{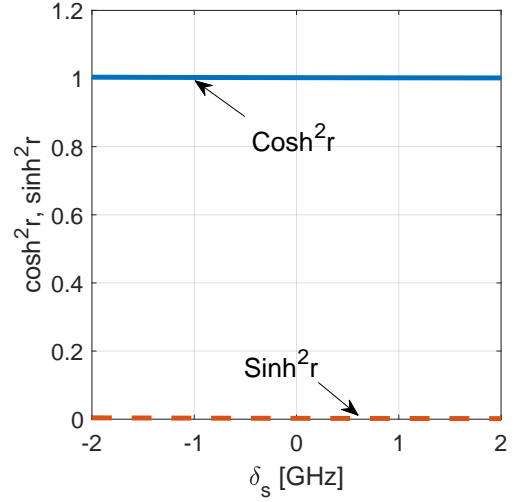}
\caption{The fractions of the diagonal modes,  $\cosh^2 r$ and $\sinh^2 r$, are plotted as a function of $\delta_s$, where the fractions are slightly changed with $\cosh^2 r\approx 1$ and $\sinh^2 r\approx 0$.}
\end{figure}

\begin{figure}
\includegraphics[width=0.8\linewidth]{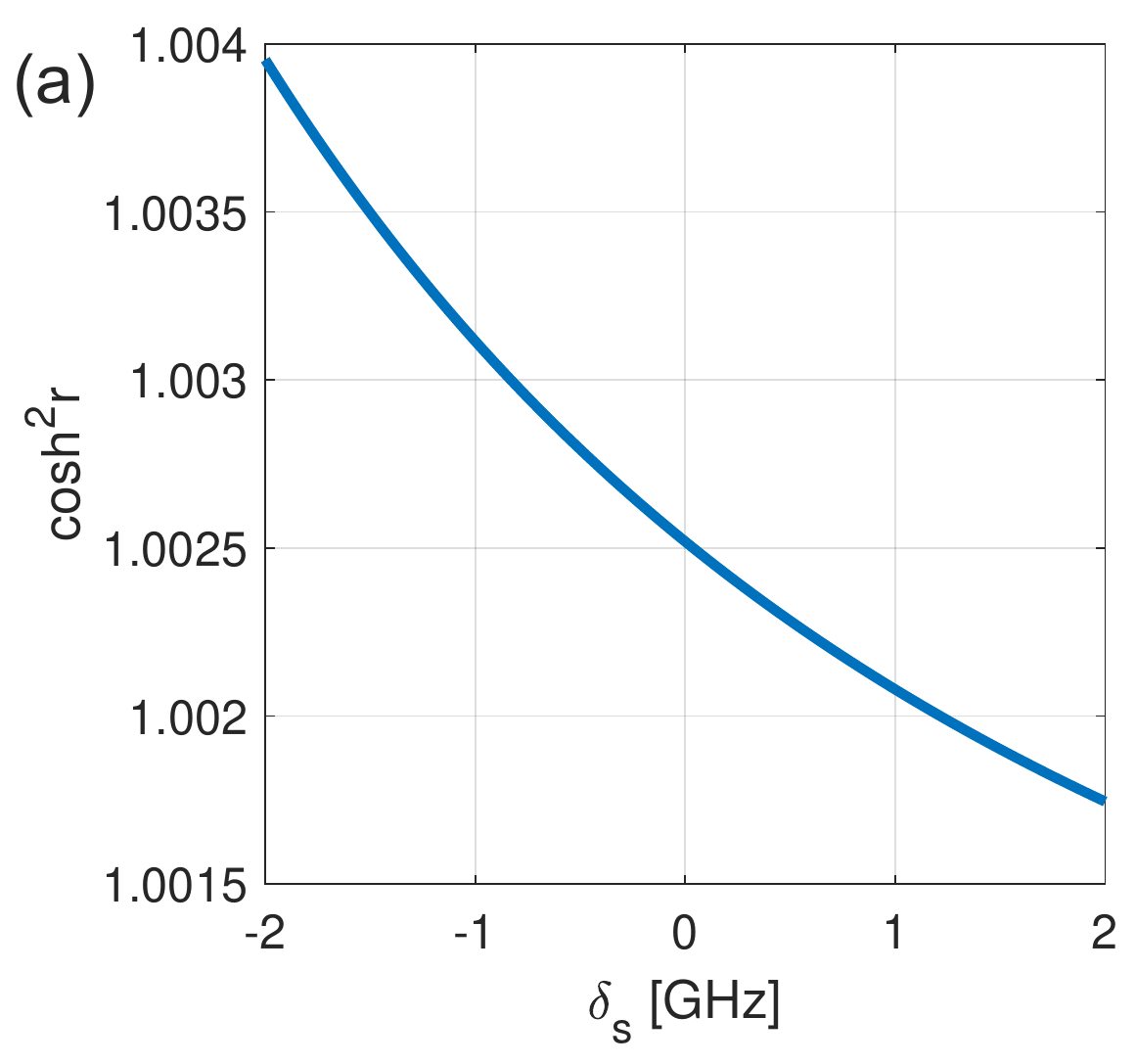}
\includegraphics[width=0.8\linewidth]{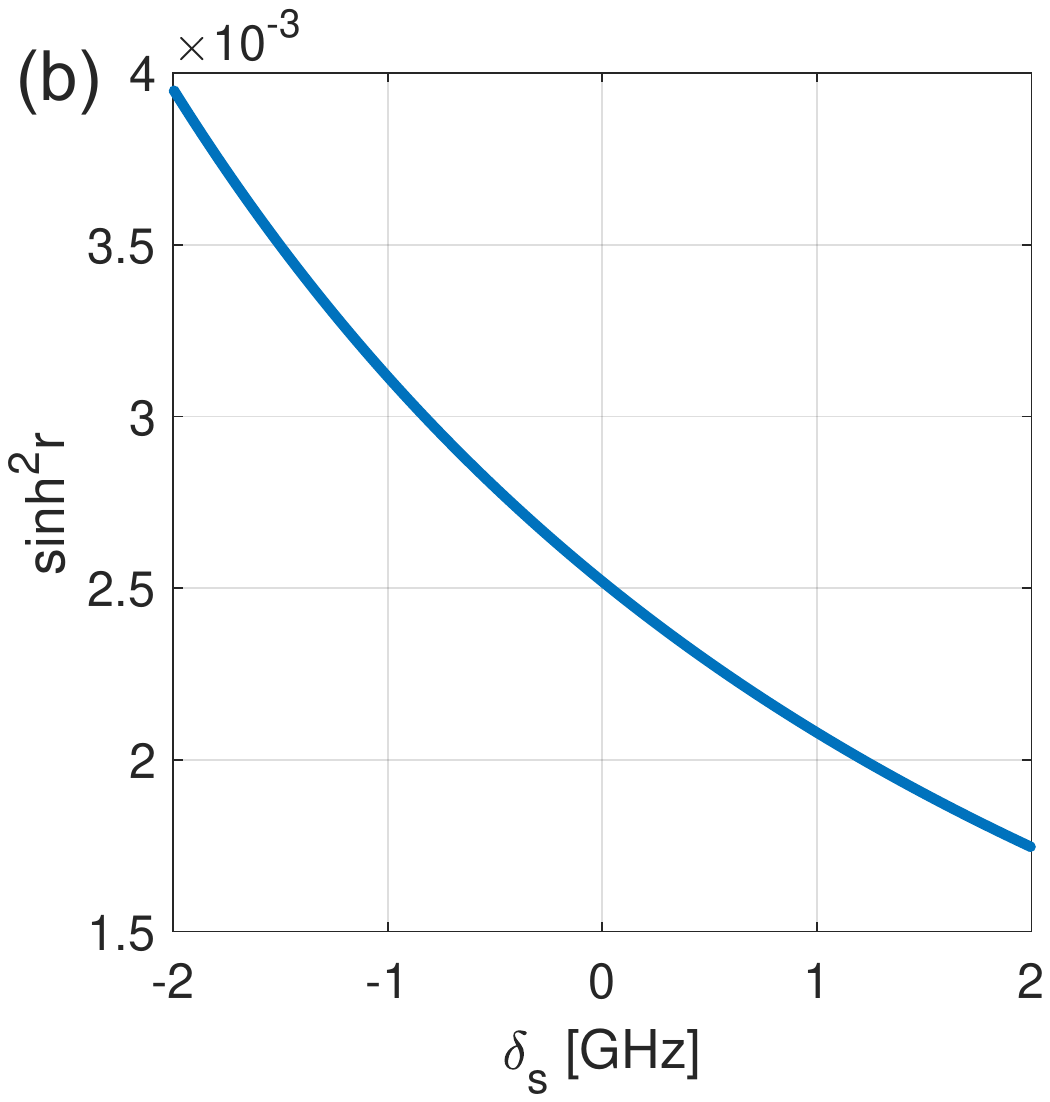}
\caption{(a) The fraction $\cosh^2 r$ of the diagonal modes is plotted as a function of $\delta_s$, where the fraction is slightly changed from one with $\cosh^2 r\approx 1$. (b) The fraction $\sinh^2 r$ of the diagonal modes is plotted as a function of $\delta_s$, where the fraction is slightly changed from zero with $\sinh^2 r\approx 0$.}
\end{figure}

The Bogoliubov transformation (\ref{TRANS}) can be achieved by defining the operator \cite{Loudon2000}
\begin{equation}
\hat{S}(r)=e^{r\left(\hat{a}_s^{\dagger}\hat{b}_s^{\dagger}-\hat{a}_s\hat{b}_s\right)},
\end{equation}
which can be written as
\begin{equation}
\hat{S}(r)=e^{\tanh r\ \hat{a}_s^{\dagger}\hat{b}_s^{\dagger}}e^{-\ln\cosh r\left(\hat{a}_s^{\dagger}\hat{a}_s+\hat{b}_s^{\dagger}\hat{b}_s+\hat{\mathbb{I}}\right)}e^{-\tanh r\ \hat{a}_s\hat{b}_s}.
\end{equation}
We apply the $\hat{S}(r)$ operator to the photon and phonon operators, by using
\begin{equation} 
\hat{\alpha}=\hat{S}^{\dagger}(r)\hat{a}_s\hat{S}(r),\ \hat{\beta}=\hat{S}^{\dagger}(r)\hat{b}_s\hat{S}(r),
\end{equation}
to yield exactly equations (\ref{TRANS}). We define the photon and phonon Fock's state $|n_{phot},n_{phon}\rangle$, where $n_{phot}$ is the number of photons and $n_{phon}$ is the number of phonons, . The vacuum state is $|vac\rangle=|0_{phot},0_{phon}\rangle$, where $\hat{a}_s|0_{phot},0_{phon}\rangle=\hat{b}_s|0_{phot},0_{phon}\rangle=0$. Next, we apply the $\hat{S}(r)$ operator on the vacuum state. We define
\begin{equation}
|r\rangle=\hat{S}(r)|vac\rangle,
\end{equation}
that yields
\begin{equation}
|r\rangle=\frac{1}{\cosh r}e^{\tanh r\ \hat{a}_s^{\dagger}\hat{b}_s^{\dagger}}|0_{phot},0_{phon}\rangle.
\end{equation}
In terms of photon and phonon Fock states we have
\begin{equation}
|r\rangle=\frac{1}{\cosh r}\sum_{n=0}^{\infty}\tanh^nr|n,n\rangle,
\end{equation}
where $|n,n\rangle=|n_{phot},n_{phon}\rangle$. The state represents entanglement between photons and phonons. Hence, if $n$ photons exist in the system then for sure exist $n$ phonons. If we measure the photonic part of the state and find $n$ photons then we are sure that $n$ phonons exist in the system. Note that in such systems one can easily perform the number of photon measurement. In the limit of $\tanh r\ll1$, we can expand the $|r\rangle$ state in terms of Fock states as
\begin{equation} \label{Ent}
|r\rangle\approx\left(|0,0\rangle+r\ |1,1\rangle+\cdots\right).
\end{equation}
Keeping the first two terms gives photon-phonon entangled state of the Bell state type \cite{Nielsen2000}. The limit of $\tanh r\ll1$ (or $r\ll1$) can be achieved at $\bar{\omega}\gg f_s$, which is consistent with the stability condition $\bar{\omega}>f_s$.

\subsection{Entangled states via anti-Stokes processes}

The anti-Stokes Hamiltonian (\ref{HAS}) in a rotating frame that oscillates with the pump frequency is given by the quadratic Hamiltonian
\begin{eqnarray} 
\hat{H}_{AS}&=&\hbar\Delta\omega_{as}\ \hat{a}_{as}^{\dagger}\hat{a}_{as}+\hbar\Omega_{as}\ \hat{b}_{as}^{\dagger}\hat{b}_{as} \nonumber \\
&+&\hbar\left( f_{as}^{\ast}\ \hat{b}_{as}^{\dagger}\hat{a}_{as}+ f_{as}\ \hat{b}_{as} \hat{a}_{as}^{\dagger}\right),
\end{eqnarray}
where $\Delta\Omega_{as}=\omega_{as}-\omega_p$, and $f_{as}=g_{as}\sqrt{\frac{n_p^{in}}{u}}$. The Hamiltonian can be diagonalized by introducing the collective operators
\begin{equation}\label{PolTr}
\hat{A}^{\pm}=X^{\pm}\ \hat{b}_{as}+Y^{\pm}\ \hat{a}_{as},
\end{equation}
which are coherent superposition of photons and phonons in the presence of the classical pump field. The coherent
mixing amplitudes are
\begin{equation}
X^{\pm}=\pm\sqrt{\frac{\Delta_{as}\mp\delta_{as}}{2\Delta_{as}}},\ \ \ Y^{\pm}=\frac{f_{as}^{\ast}}{\sqrt{2\Delta_{as}(\Delta_{as}\mp\delta_{as})}},
\end{equation}
where
\begin{equation}
\Delta_{as}=\sqrt{\delta_{as}^2+|f_{as}|^2},\ \ \ \delta_{as}=\frac{\Delta\omega_{as}-\Omega_{as}}{2}.
\end{equation}
We have the normalization condition $|X^{\pm}|^2+|Y^{\pm}|^2=1$. The diagonal Hamiltonian reads
\begin{equation}
\hat{H}_{AS}=\sum_{\mu=\pm}\hbar\Omega_{\mu}\ \hat{A}^{\mu\dagger}\hat{A}^{\mu},
\end{equation}
with the diagonal dispersions
\begin{equation}
\Omega_{\pm}=\frac{\Delta\omega_{as}+\Omega_{as}}{2}\pm \Delta_{as}.
\end{equation}
The collective operators $\hat{A}^{\pm\dagger}$ create photon-phonon entangled states out of the vacuum state  $|vac\rangle$ with energies $\hbar\Omega_{\pm}$. Such coherent states usually are termed polaritons \cite{Zoubi2005,Zoubi2007}.

We apply the collective operators to the un-entangled state $|n_{phot},m_{phon}\rangle$ to get the entangled state
\begin{eqnarray}
\hat{A}^{\pm\dagger}|n_{phot},m_{phon}\rangle&=&Y^{\pm\ast}\ \sqrt{n_{phot}+1}|n_{phot}+1,m_{phon}\rangle \nonumber \\
&+&X^{\pm\ast}\ \sqrt{n_{phon}+1}|n_{phot},m_{phon}+1\rangle. \nonumber \\
\end{eqnarray}
For example the vacuum state $|0_{phot},0_{phon}\rangle$ casts into the entangled state $\left(Y^{\pm\ast}\ |1_{phot},0_{phon}\rangle+X^{\pm\ast}\ |0_{phot},1_{phon}\rangle\right)$. At resonance where $\Delta\omega_{as}=\Omega_{as}$ we have $\delta_{as}=0$, and for real $f_{as}$ we get $\Delta_{as}=f_{as}$, hence $X^{\pm}=\pm\frac{1}{\sqrt{2}}$ and $Y^{\pm}=\frac{1}{\sqrt{2}}$. The entangled state is part of the Bell states  $\frac{1}{\sqrt{2}}\left(|1_{phot},0_{phon}\rangle\pm|0_{phot},1_{phon}\rangle\right)$. The measurement of the photonic part of the entangled state gives information about the phonon number in the system.

For typical silicon nanowires we use the previous numbers. The photon-phonon coupling parameter, $g_{as}=1$ MHz, and for the internal-external coupling parameter at the multiplexers, $u=1$ MHz, and the average number of the incident pump photons is taken to be of the order of $10^{12}$ per second, where for the effective photon-phonon coupling parameter we get $f_{as}=1$ GHz. For the pump photon we use $\omega_p=10^{15}$ Hz, and for the phonon we have $\Omega_{as}=10$ GHz. The diagonal frequencies are plotted in figure (10). The photon and phonon are coherently mixed (in the presence of the pump field) to form the entangled states. The splitting between the two diagonal states at $\delta_{as}=0$ equals to $2f_{as}$. The photon and photon fractions in the diagonal states are plotted in figures (11). Around $\delta_{as}=0$ the diagonal state is half photon and half phonon, that is $|X^{\pm}|^2=|Y^{\pm}|^2=1/2$. Far from the intersection point, the lower branch becomes phononic at negative values of $\delta_{as}$ and photonic at positive values, and vise versa  the upper branch becomes photonic at negative values of $\delta_{as}$ and phononic at positive values.

Here, the phonon damping rate, $\Gamma=1$ MHz \cite{Kharel2016}, is of the order of the photon-phonon coupling parameter, $g_{as}$. But the phonon damping rate is much smaller than the effective photon-phonon coupling parameter $f_{as}$. Then we are in the strong coupling regime, and the appearance of the diagonal modes is achievable. As mentioned before in order to minimize the influence of the thermal phonons at 10 GHz frequency one needs to cool the nanowire down to about 10 mK.

\begin{figure}
\includegraphics[width=0.8\linewidth]{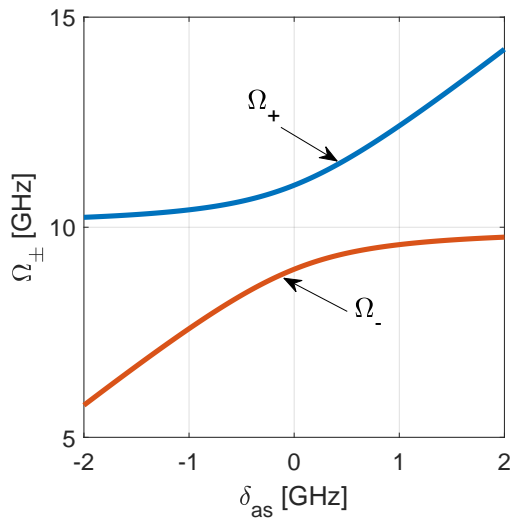}
\caption{The diagonal modes, $\Omega_{\pm}$ are plotted as a function of the detuning $\delta_s$. The splitting between the two diagonal modes is $2f_{as}$ at $\delta_{as}=0$.}
\end{figure}

\begin{figure}
\includegraphics[width=0.8\linewidth]{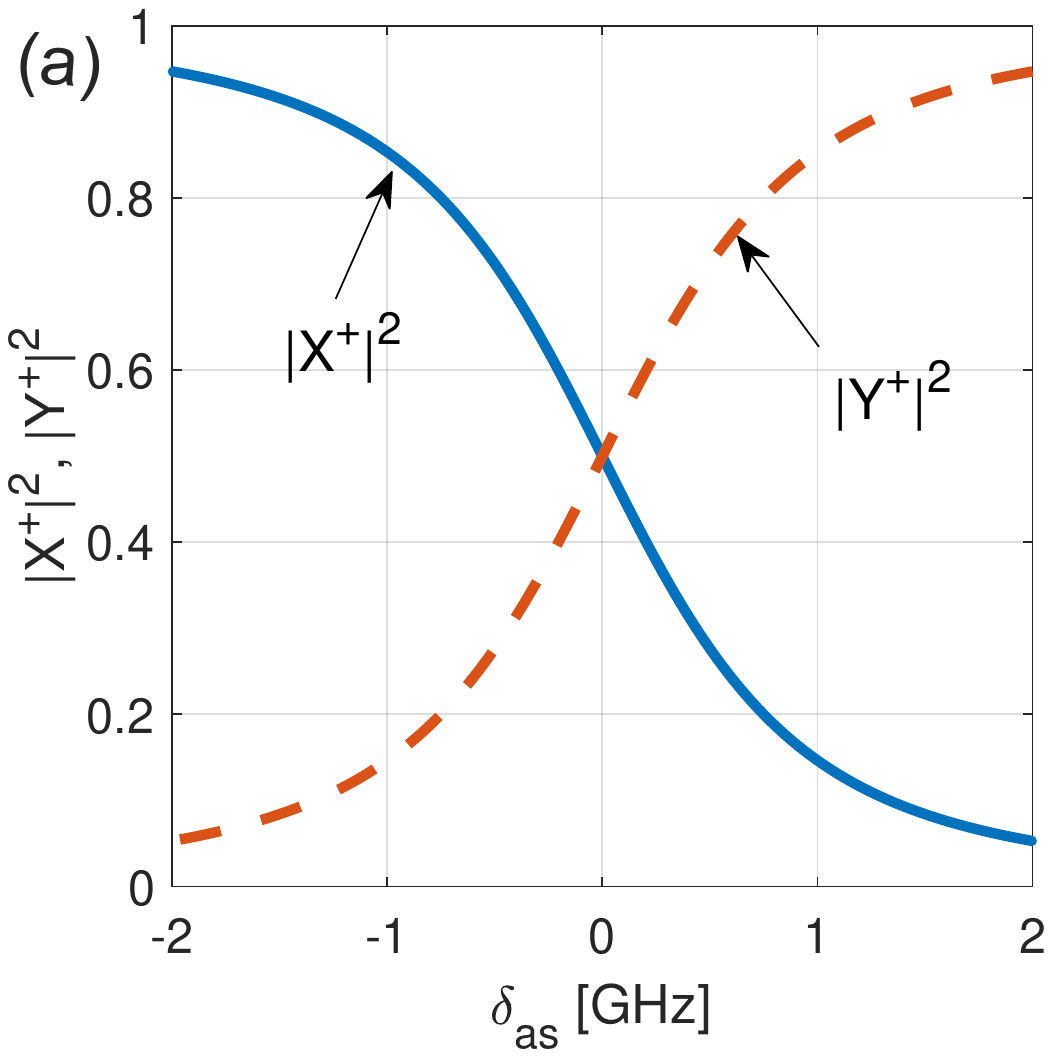}
\includegraphics[width=0.8\linewidth]{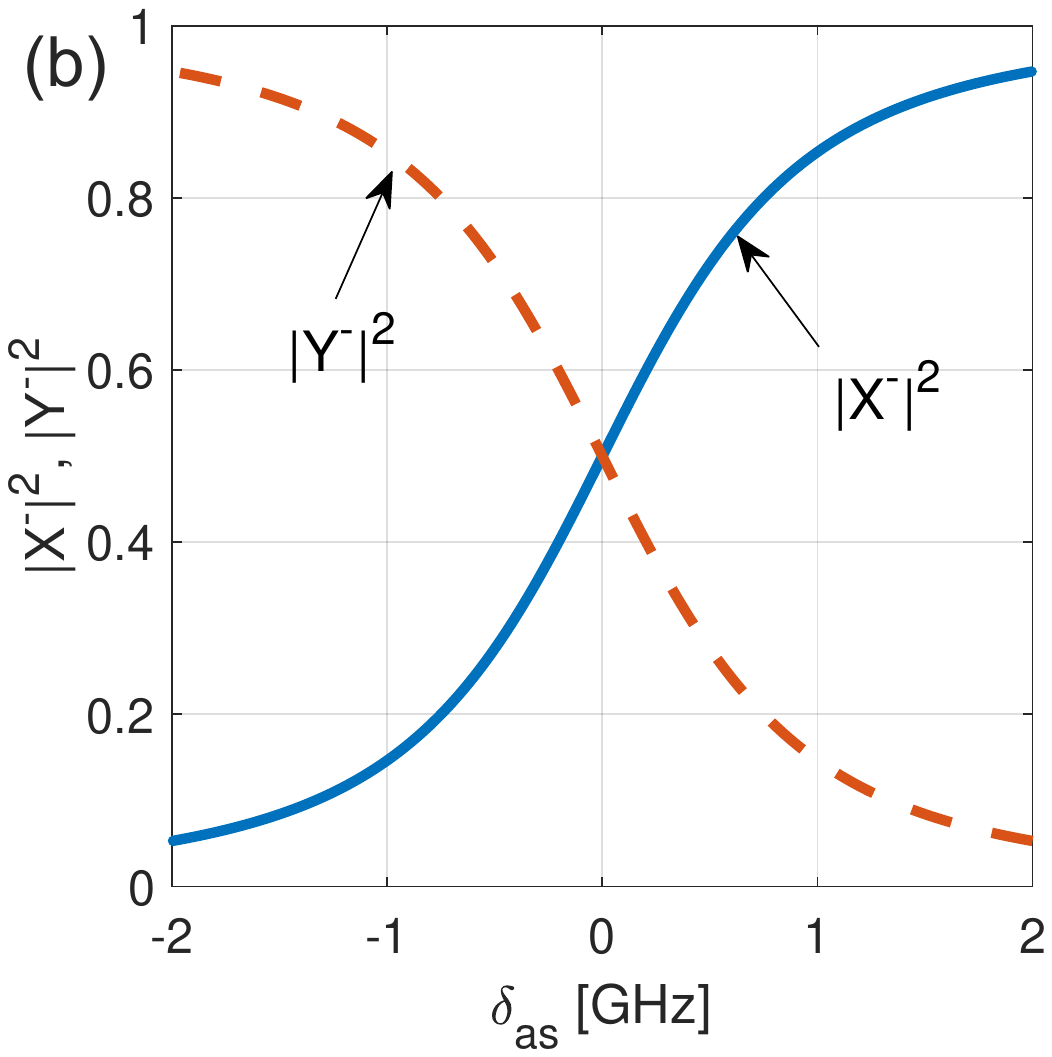}
\caption{(a) The phononic fraction $|X^{+}|^2$ and the photonic fraction $|Y^{+}|^2$ of the upper diagonal mode $(+)$ is plotted as a function of the detuning $\delta_{as}$. (b) The phononic fraction $|X^{-}|^2$ and the photonic fraction $|Y^{-}|^2$ of the lower diagonal mode $(-)$ is plotted as a function of the detuning $\delta_{as}$. In both cases the fractions are half around the point of $\delta_{as}=0$, and become photonic or phononic far from this point.}
\end{figure}

\section{Conclusions}

Nanoscale waveguides provide a tool for manipulating photons and phonons and for controlling their interactions, mainly due to the new mechanism of radiation pressure that becomes dominant once the light wavelength is smaller than the waveguide dimension. This fact transformed Brillouin's scattering from an obstacle for efficient communications into a useful mean for information processing. Electronic devices made of silicon are the basic elements for digital computations, and nanowires made of silicon extend the use of silicon into the field of photonics and convert them to be a strong candidate for quantum information processing involving photons. Photonic silicon components can be easily integrated into on-chip network, which is a big step toward all optical platform. Electronic and photonic devices made of silicon can be combined together on-chip in order to perform processing and communication at the same platform.

In the present paper we introduced a new mechanism for the possibility of the formation of photon-phonon entangled states. To this end we exploited Brillouin scattering inside nanoscale waveguides made of silicon that yields strong photon-phonon coupling. Photons of a strong pump field are stimulated to scatter either into lower or higher frequency photons in the existence of a weak probe field. In the Stokes process the scattering is stimulated into lower frequency photons by the emission of phonons, while in the anti-Stokes process the scattering is into higher frequency photons by the absorption of phonons. The pump and probe photons belong to distinct spatial modes and the Stokes and anti-Stokes phonons differ by their wavenumber. Hence the Stokes and anti-Stokes processes decouple and can be treated separately. The nonlinear Brillouin's Hamiltonian, that includes three operators, is converted into a linear Hamiltonian by taking the strong pump field to be a classical one. The linear Hamiltonian is diagonalized to yield collective photon-phonon states which are coherent superposition of photon-phonon states. The states show entanglement properties and we investigated the possibility of getting two types of Stokes and anti-Stokes photon-phonon entangled states.

In the paper we proposed a silicon nanowire as a physical component for the generation of entangled states between photons and phonons. The idea can be realized at the current experimental set up and is of importance for quantum information processing and quantum communications. The measurement of the photonic part of the entangled state provides a source at the level of single phonons in demand. The photons can travel a long distance away from the nanowire and stay entangled with the phonons. Moreover, such photons can swap their entanglement to other phonons at a distance nanowire and lead to entangled phonons among nanowires separated by a large distance. We plan to perform more studies about quantum information processing and quantum communications involving photons and phonons in a network of nanoscale waveguides.

\end{document}